\documentclass[twocolumn,showpacs,preprintnumbers,amsmath,amssymb]{revtex4}
\usepackage{graphicx}
\usepackage{dcolumn}
\usepackage{bm}

\begin{document}

\title{Strain-controlled valley and spin separation in silicene heterojunctions}
\author{Yuan Li$^{1,2}$}
\email{liyuan@hdu.edu.cn}
\author{H. B. Zhu$^{1}$}
\author{G. Q. Wang$^{1}$}
\author{Y. Z. Peng$^{1}$}
\author{J. R. Xu$^1$}
\author{Z. H. Qian$^{2}$}
\author{R. Bai$^{2}$}
\author{G. H. Zhou$^3$}
\author{C. Yesilyurt$^4$}
\author{Z. B. Siu$^4$}
\author{M. B. A. Jalil$^{4}$}
\email{elembaj@nus.edu.sg}

\affiliation{$^1$Department of Physics, Hangzhou Dianzi University, Hangzhou, Zhejiang 310018,
China}
\affiliation{$^2$Center for Integrated Spintronic Devices (CISD), Hangzhou Dianzi University, Hangzhou, Zhejiang 310018,
China}
\affiliation{$^3$Department of Physics and Key
Laboratory for Low-Dimensional Quantum Structures and Manipulation
(Ministry of Education), Hunan Normal University, Changsha 410081,
China}
\affiliation{$^4$ Computational Nanoelectronics and Nano-device Laboratory, Electrical and Computer
Engineering Department, National University of Singapore, 4 Engineering Drive 3,
Singapore 117576, Singapore}

\begin{abstract}
We adopt the tight-binding mode-matching method to study the strain effect on silicene heterojunctions.
It is found that valley and spin-dependent separation of electrons cannot be achieved by the electric field only. When a strain
and an electric field are simultaneously applied to the central scattering region, not only are the electrons of valleys K and K'
separated into two distinct transmission lobes in opposite transverse directions,
but the up-spin and down-spin electrons will also move in the two opposite transverse directions.
Therefore, one can realize an effective modulation of valley and
spin-dependent transport by changing the amplitude and the stretch direction of the strain. The phenomenon
of the strain-induced valley and spin deflection can be exploited for silicene-based valleytronics devices.
\end{abstract}
\date{\today}
\maketitle

\section{Introduction}
Silicene, a low-buckled monolayer-honeycomb lattice of silicon atoms, has
been synthesized on metal surfaces~\cite{Aufray, Padova, Lalmi} and attracted extensive attention both
theoretically~\cite{Fagan,Cahangirov} and experimentally~\cite{Vogt,Chen} recently.
Its low-buckled structure supports a relatively large spin-orbit coupling (SOC) and
a sizable gap of $1.55 \mathrm{meV}$ at the Dirac points K and K'~\cite{Guzman-Verri,Liu}.
The band gap of silicene can be modulated by applying a perpendicular electric field, thus inducing
a topological phase transition as the electric field increases~\cite{Drummond,Ezawa2,An}.
 The compatibility of silicene with silicon-based technology motivates many studies of interesting effects,
such as the spin- and valley-Hall effects~\cite{Liu2,Tabert,Missault}, the quantum anomalous Hall effect~\cite{Ezawa,Pan},
valley-spin coupling~\cite{Stille,Yesilyurt}.

The existence of the spin-valley coupling makes silicene be a candidate for valleytronics. However,
the SOC is  weak compared with transition metal dichalcogenides (TMDs).
The interplay of spin, valley and Berry phase related physics in TMDs, such as MoS$_2$ and WSe$_2$,
can result in a valley-dependence spin Hall effect~\cite{Shan,Xiao}. Compared with TMDs, it seems that silicene is not
suitable for switching operations in valleytronics devices due to the weak SOC. Thus it is desirable to
create a large band gap and SOC in silicene systems so as to catch up with TMDs in valleytronics. Recently,
first-principles calculations show that the energy band can be significantly modulated by applying a strain
in silicene systems~\cite{Wang,Qu}. The strain-induced band gap of silicene structures can reach
the maximum value of $0.08 \mathrm{eV}$~\cite{Zhao2}. Obviously, the strain can induce a large band gap, which is
comparable to that of TMDs and suitable for switching operations in valleytronics devices. Experimentally,
one can realize a controllable strain in silicene via deposition onto stretchable substrates,
similar to the strain effect in MoS$_2$~\cite{Gomez}, or by exerting an external mechanical force.

However, the effect of the strain on the valley and spin separation in silicene systems
has not been discussed previously. In this paper, we adopt the tight-binding mode-matching method and
propose an efficient way to separate the Dirac fermions
of different valleys, thus create a distinct spin separation by utilizing the strain and the electric field
in silicene systems. Our results show that the valley- and spin-dependent electrons cannot be dispersed only by the
electric field. Combining the strain and the electric field, one can realize an effective modulation
of valley and spin-dependent transport by changing the amplitude or the stretch direction of the strain,
without the need for ferromagnetic materials or magnetic fields.
This phenomenon provides a novel route to effectively modulate the valley and spin polarizations
of the silicene devices by utilizing the strain and the electric fields.
\begin{figure}[t]
\includegraphics[width=9cm]{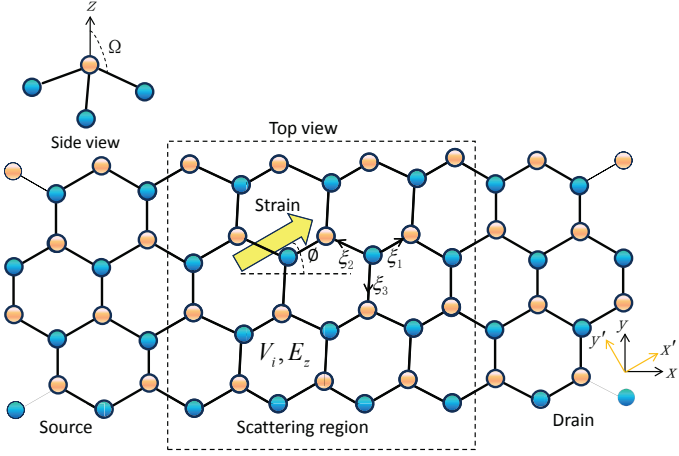}
\caption{\label{fig:device} Schematic of the silicene heterojunction with an uniaxial strain, electric field and voltage potential in the central scattering region. The zigzag direction of the honeycomb lattice (x-y plane) is always parallel to the axis $x$, the tension is applied along the angle $\phi$ relative to the axis $x$, and the angle $\Omega$ is defined as between the Si-Si bond and the z direction normal to the plane.}
\end{figure}

Comparing with the tight-binding model, the Dirac theory is an effective approach which can only serve as a starting point for theoretical studies of transport in silicene. It has the advantage of yielding analytical results which capture the basic physical insights for certain problems, especially those with simplified system geometries. However, for a general consideration, e.g., for complicated geometries, or for spatially dependent variation in the lattice configuration (e.g. due to strain or defects), the low-energy Dirac Hamiltonian is not readily available, and one has to resort to the more general tight-binding model. Furthermore, the tight-binding model would automatically include higher-order terms and the contribution of both the K and K' valleys, and allow for the complete band information to be captured (even for spatially varying systems). Furthermore, the effects of leads and other interactions (impurity scattering, etc.) can be included systematically in the tight-binding model combined with the non-equilibrium Green's function approach (NEGF) and the mode-matching method. The tight-binding NEGF formalisms form the basis of quantum transport modeling of nanoscale devices~\cite{Datta}. It can deal with a wide range of conductors, composed of a scattering region and external leads, under the application of a bias. Therefore, it is important to develop and demonstrate the use of the tight-binding NEGF technique and the mode-matching method in this paper, due to its more general application than that of the effective Dirac Hamiltonian.

The paper is organized as follows. In Sec. \ref{sec:Model}, we introduce the system under consideration, i.e., a silicene heterojunction under the influence of strain and external electric field applied to the central scattering region. We then calculate the strain-modulated hopping parameters based on the Slater-Koster framework, and analyze the dispersion relations. In Sec. \ref{sec:transport}, we employ the mode-matching method to investigate the spin and valley-dependent angular transmissions. In Sec. \ref{sec:Results}, the combined effects of the strain and the electric field on the valley and spin separation are analyzed and discussed. A summary is given in Sec. \ref{sec:Conclusions}.

\section{MODEL AND DISPERSION RELATIONS}\label{sec:Model}
We consider a low-buckled silicene sheet with zigzag direction along the axis $x$,
in which the angle $\Omega$ describes the amplitude of the buckling with lattice constant
being $a=3.86{\AA}$.
In the central scattering region, the silicene sheet is stretched (or compressed) along the angle $\phi$
relative to the axis $x$, as shown in Fig.~\ref{fig:device}. Note that we assume there exists no strain outside the central scattering region.
The silicene sheet can be described by the four-band second-nearest-neighbor tight-binding model~\cite{Liu,Ezawa}
\begin{eqnarray}\label{eq:parameter}
H&=&\sum_{i\alpha}V_ic^\dag_{i\alpha} c_{i\alpha}+i\frac{t_{so}(\vec{\xi})}{3\sqrt{3}}\sum_{\langle\langle i,j\rangle\rangle \alpha\beta}\nu_{ij}c^\dag_{i\alpha}\sigma_{\alpha\beta}^z c_{j\beta}\nonumber \\
&&-i\frac{2}{3}t_{R2}(\vec{\xi})\sum_{\langle\langle i,j\rangle\rangle \alpha\beta}\mu_{i}c^\dag_{i\alpha}(\vec{\sigma}\times \hat{\mathbf{d}}_{ij})^z_{\alpha\beta}c_{j\beta}\nonumber\\
&&-t(\vec{\xi})\sum_{\langle i,j\rangle\alpha}c^\dag_{i\alpha} c_{j\alpha}-\sum_{i\alpha}\mu_ia_zE_zc^\dag_{i\alpha} c_{i\alpha},
\end{eqnarray}

\begin{figure*}[t]
\includegraphics[width=8cm]{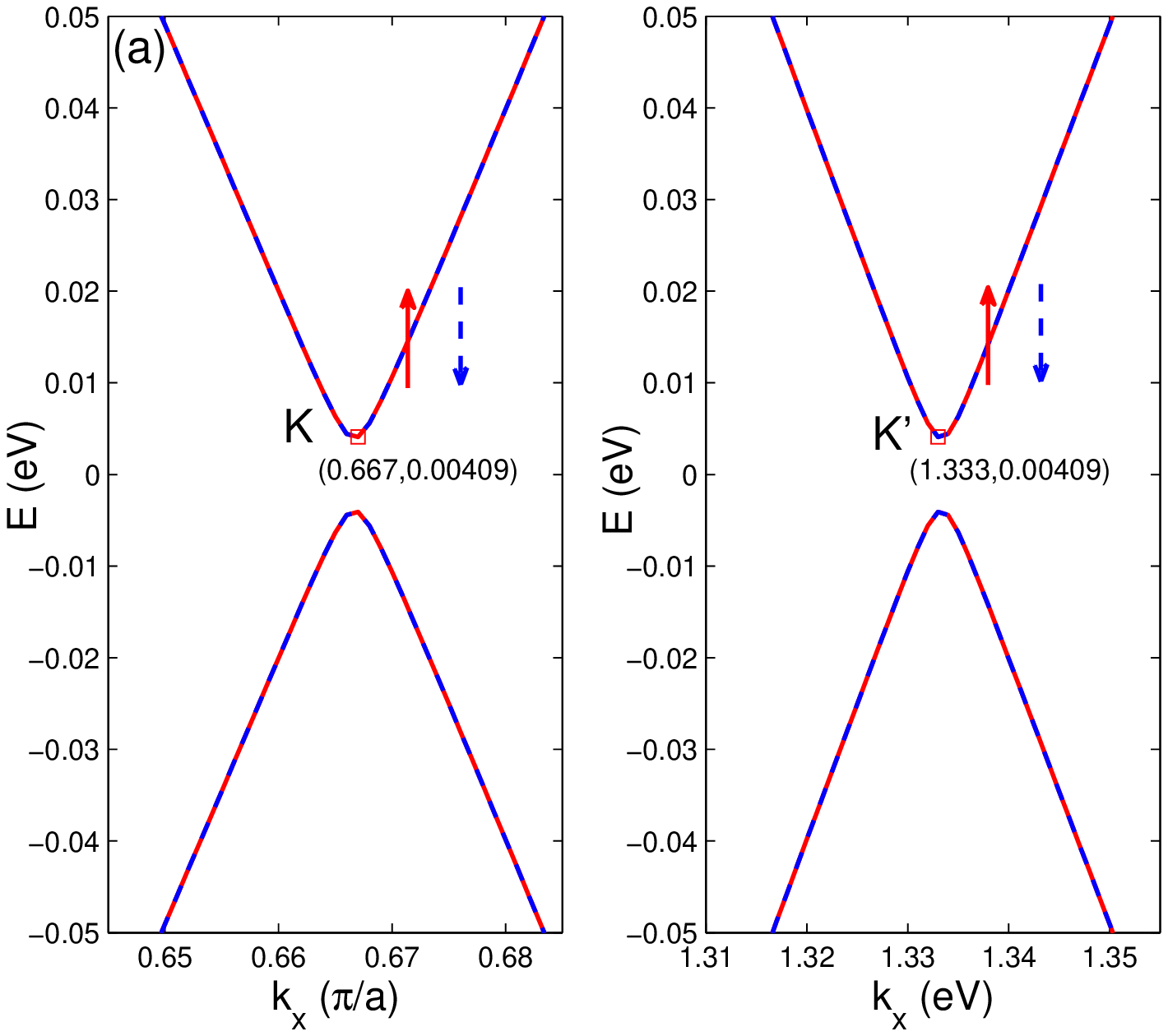}
\includegraphics[width=8cm]{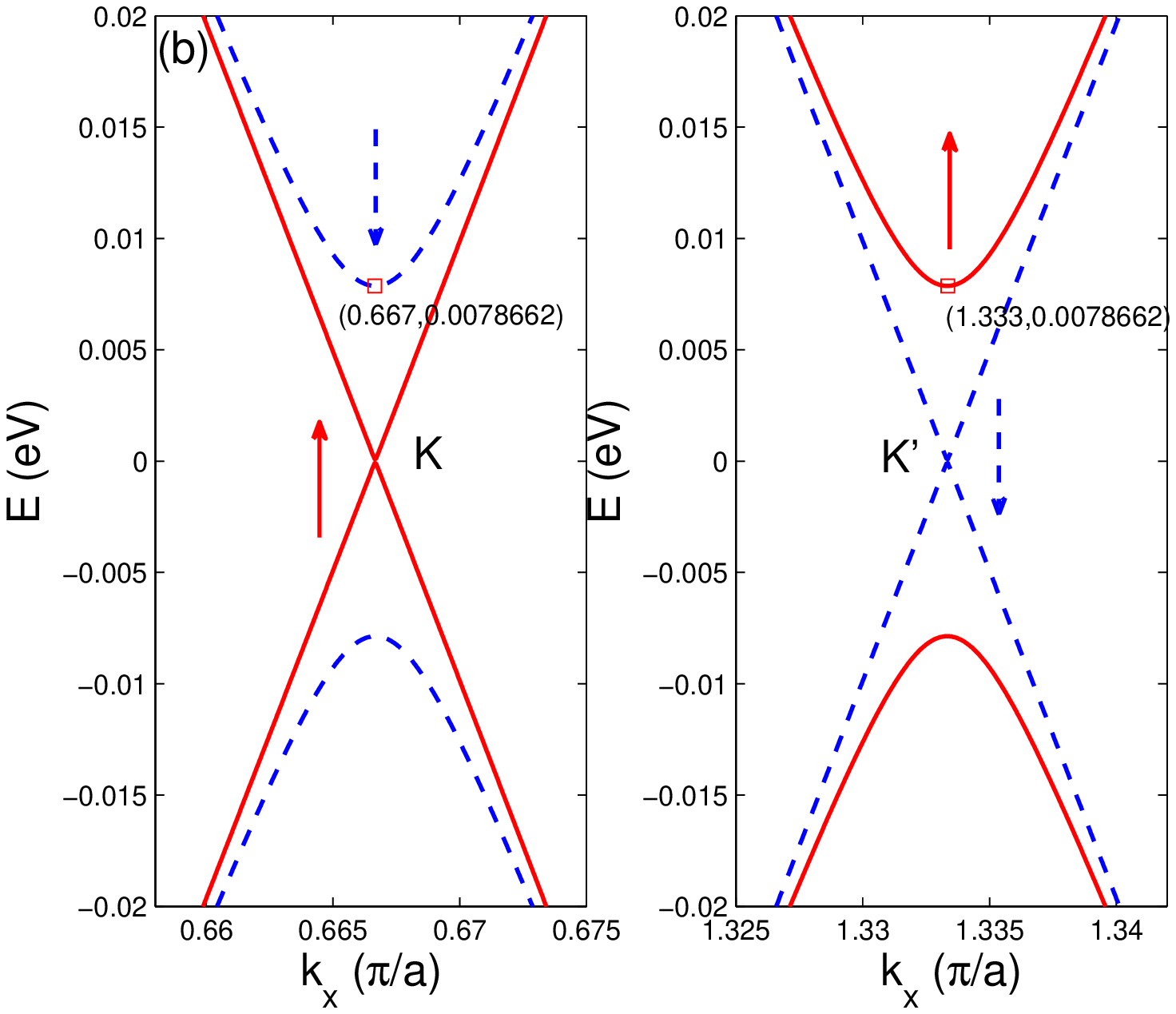}
\includegraphics[width=8cm]{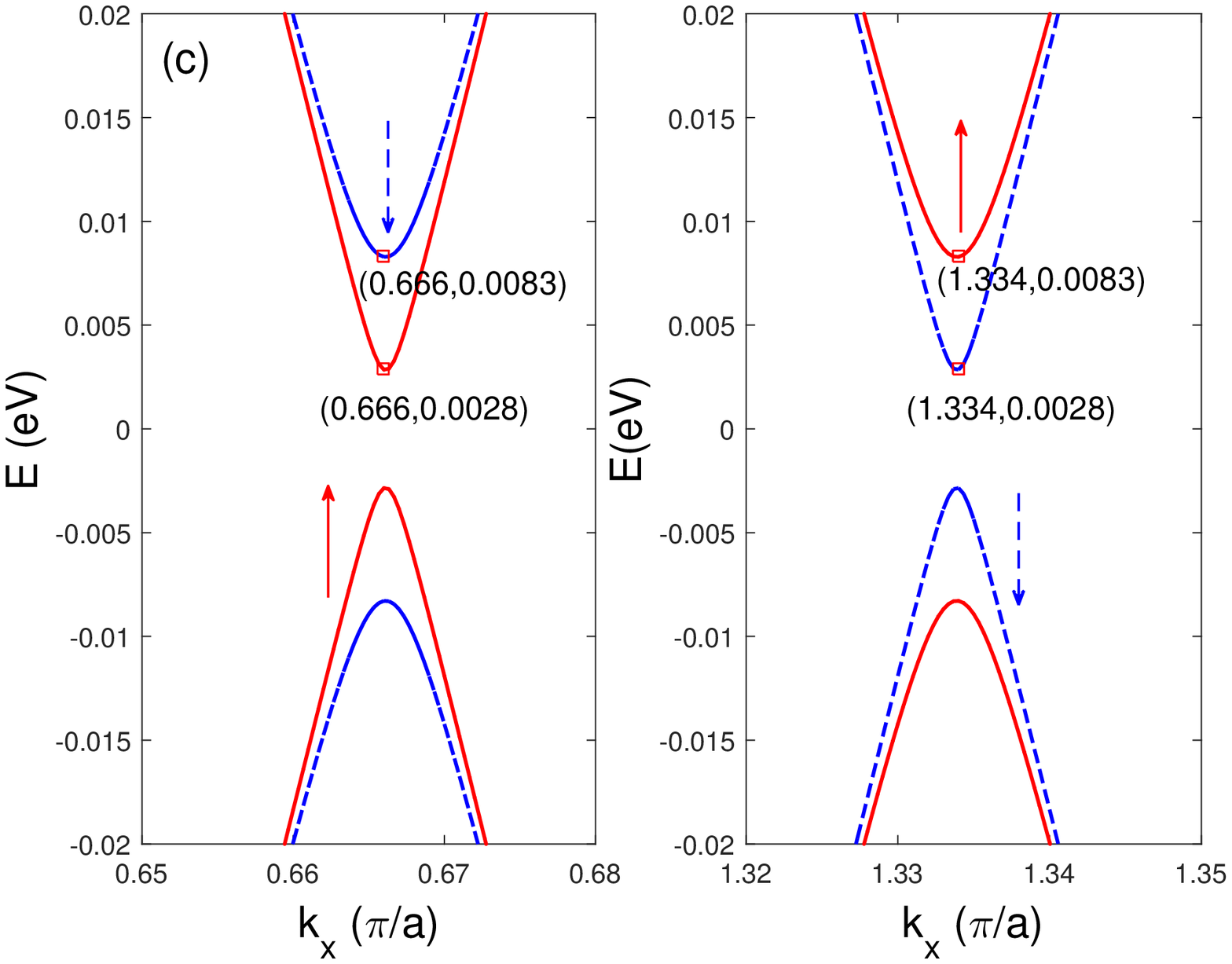}
\includegraphics[width=8cm]{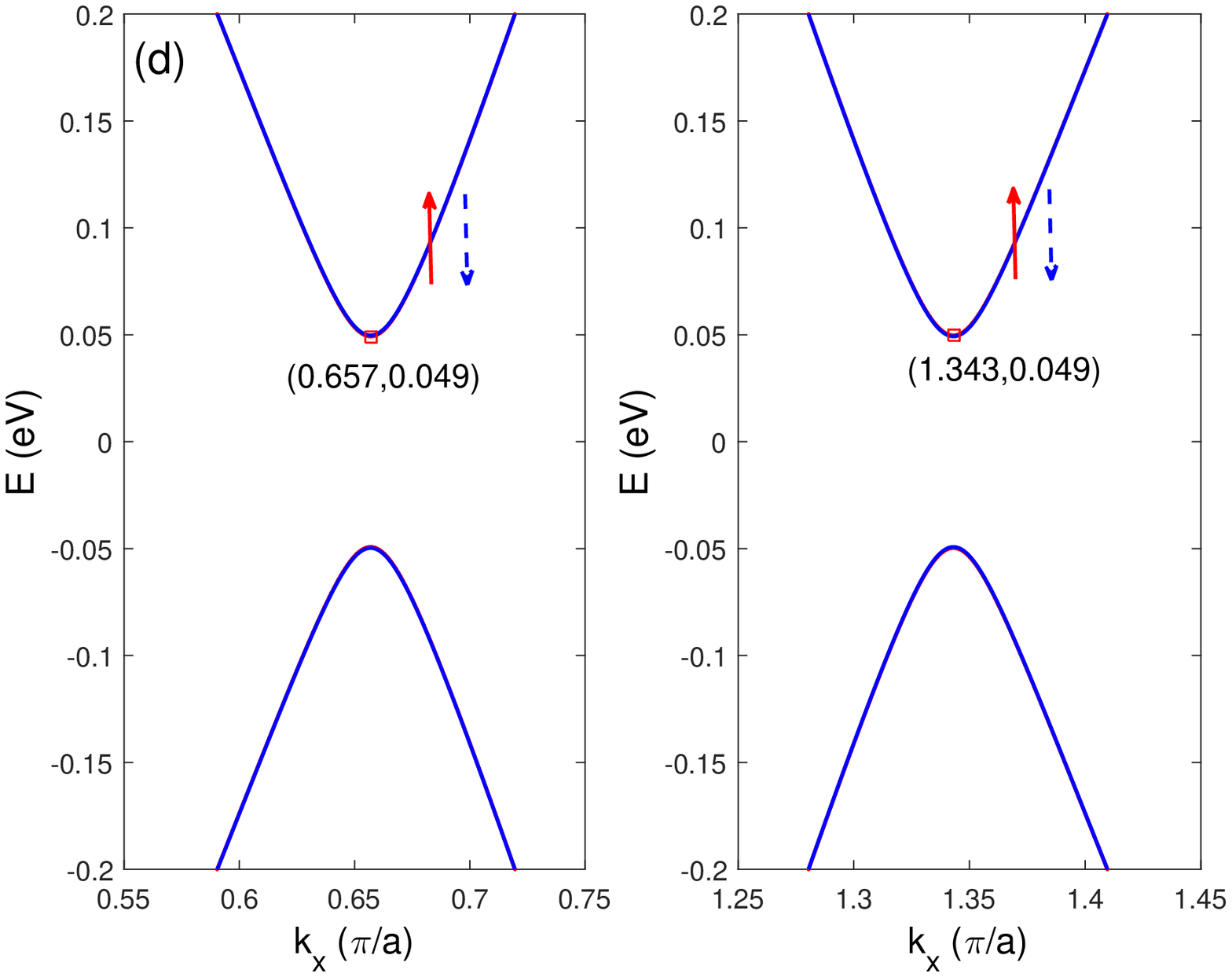}
\caption{\label{fig:dispersion}  The dispersion relations are plotted as a function of the wave vector $k_x$ for
(a) $E_z=0$, $\epsilon_0=0$, the energy difference is $\triangle E=8.2 \mathrm{meV}$,
(b)$E_z=16.96 \mathrm{meV}{\AA}$, $\epsilon_0=0$,
(c)$E_z=16.96 \mathrm{meV}{\AA}$, $\epsilon_0=0.005 $,
(d)$E_z=16.96 \mathrm{meV}{\AA}$, $\epsilon_0=0.05$.
The other parameters are $\nu_{\parallel}=0.25$, $\nu_{\perp}=2.5$, $\phi=30^\circ$ and $k_y=0$.
The red and blue arrows refers to the spin indexes.}
\end{figure*}
where $c^\dag_{i\alpha} (c_{i\alpha})$ refers to the creation (annihilation) operator with spin index $\alpha$ at site $i$,
and $\langle i,j \rangle/\langle\langle i,j\rangle \rangle$ run over all the nearest or next-nearest neighbor hopping sites.
The first term is the on-site potential energy, the second term denotes the effective spin-orbit coupling with the hopping parameter $t_{so}(\vec{\xi})$, where $\vec{\sigma}=(\sigma_x,\sigma_y,\sigma_z)$ are the spin Pauli matrix operators,
and $\nu_{ij}=\pm 1$ for the anticlockwise (clockwise)
hopping between the next-nearest-neighboring sites with respect to the positive $z$ axis.
The third term represents the Rashba spin-orbit coupling with $\mu_i=\pm1$ for the A(B) site,
where $\hat{\mathbf{d}}_{ij}=\mathbf{d}_{ij}/|\mathbf{d}_{ij}|$ refers to the unit vector connecting the two
next-nearest-neighboring sites.
The fourth term is the nearest-neighbor hopping with the transfer energy $t(\vec{\xi})$,
where the vector $\vec{\xi}$ is adopted to describe the elastic response for which deformations are
affine~\cite{Pereira2}.
The fifth term describes the contribution of the staggered sublattice potential,
with $2a_z=0.46 {\AA}$ being the distance of the two sublattice planes.
The relaxed equilibrium values for the hopping parameters are
$t^0(\vec{\xi})\approx 1.09 \mathrm{eV}$, $t_{so}^0(\vec{\xi})\approx 3.9 \mathrm{meV}$ and
$t_{R2}^0(\vec{\xi})\approx 0.7 \mathrm{meV}$~\cite{Liu}.

\begin{figure*}[t]
\includegraphics[width=8cm]{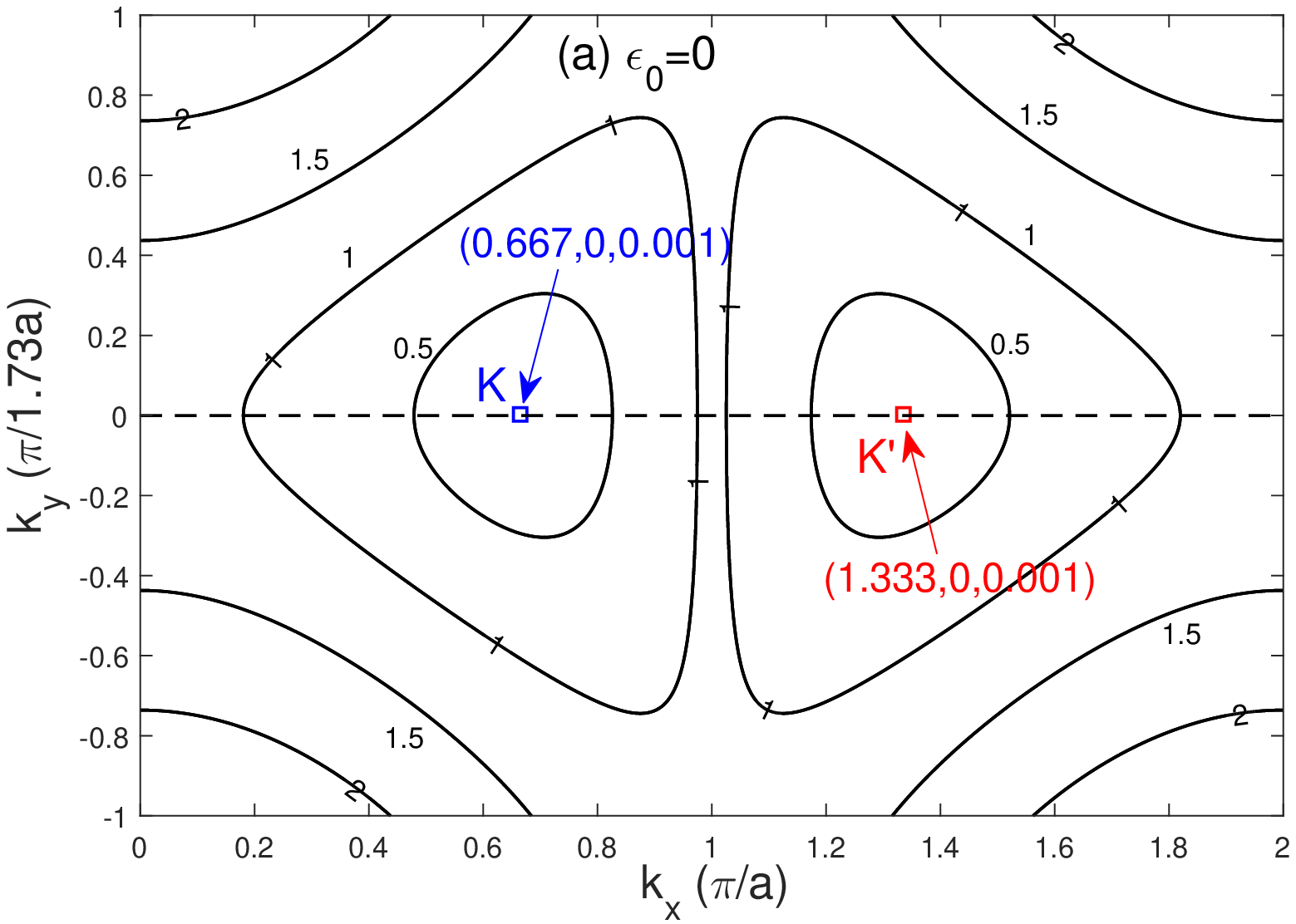}
\includegraphics[width=8cm]{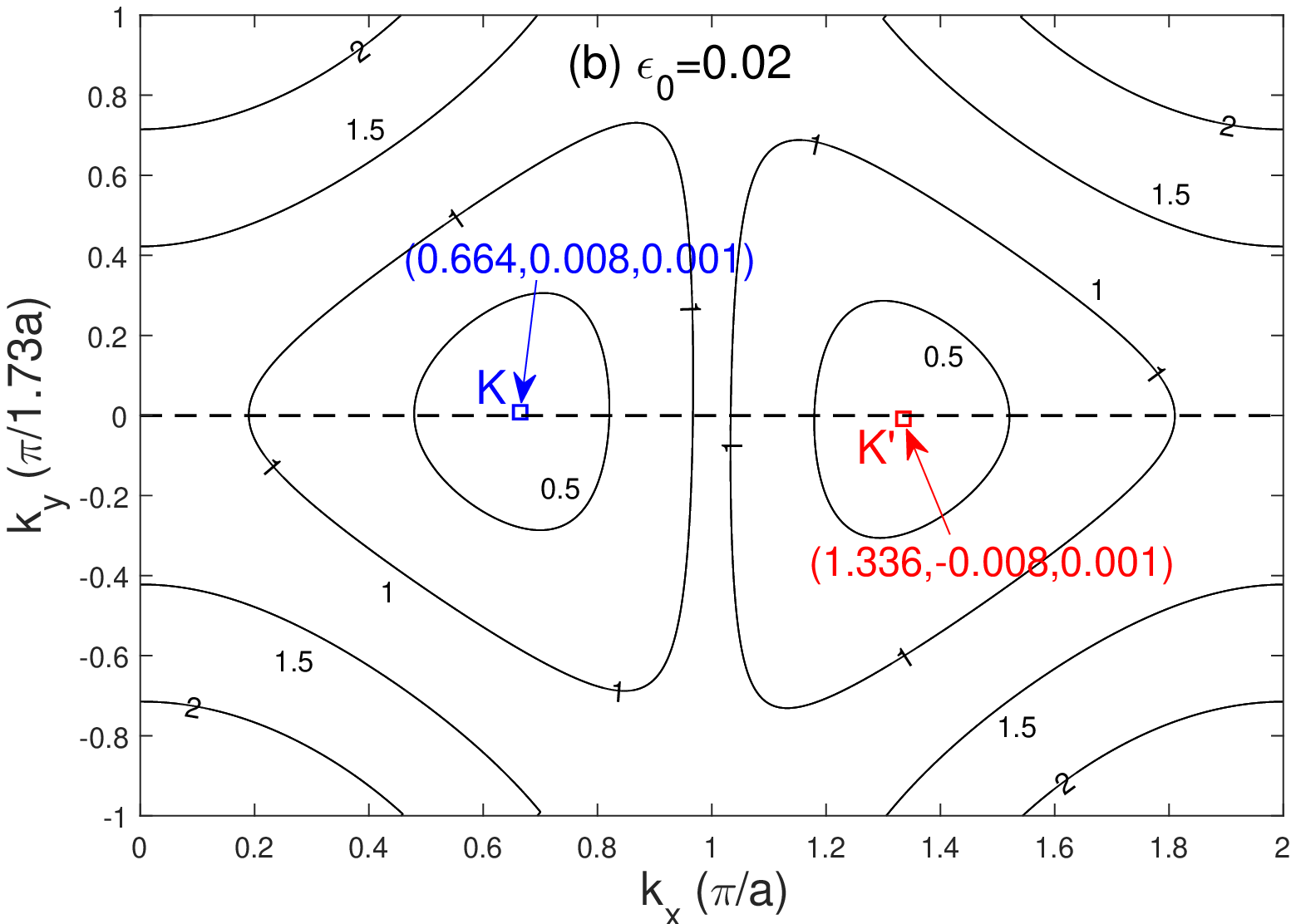}
\includegraphics[width=8cm]{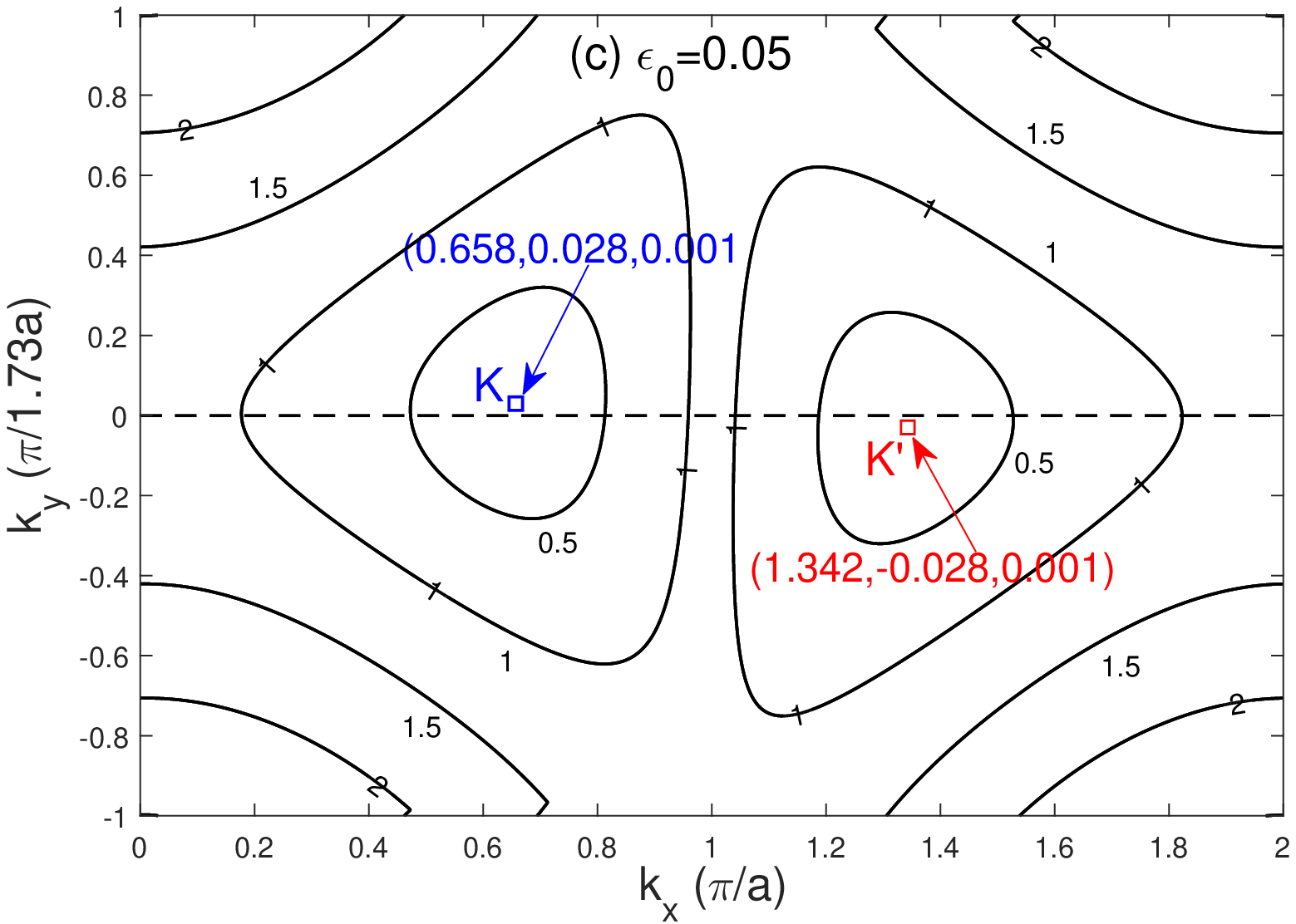}
\includegraphics[width=8cm]{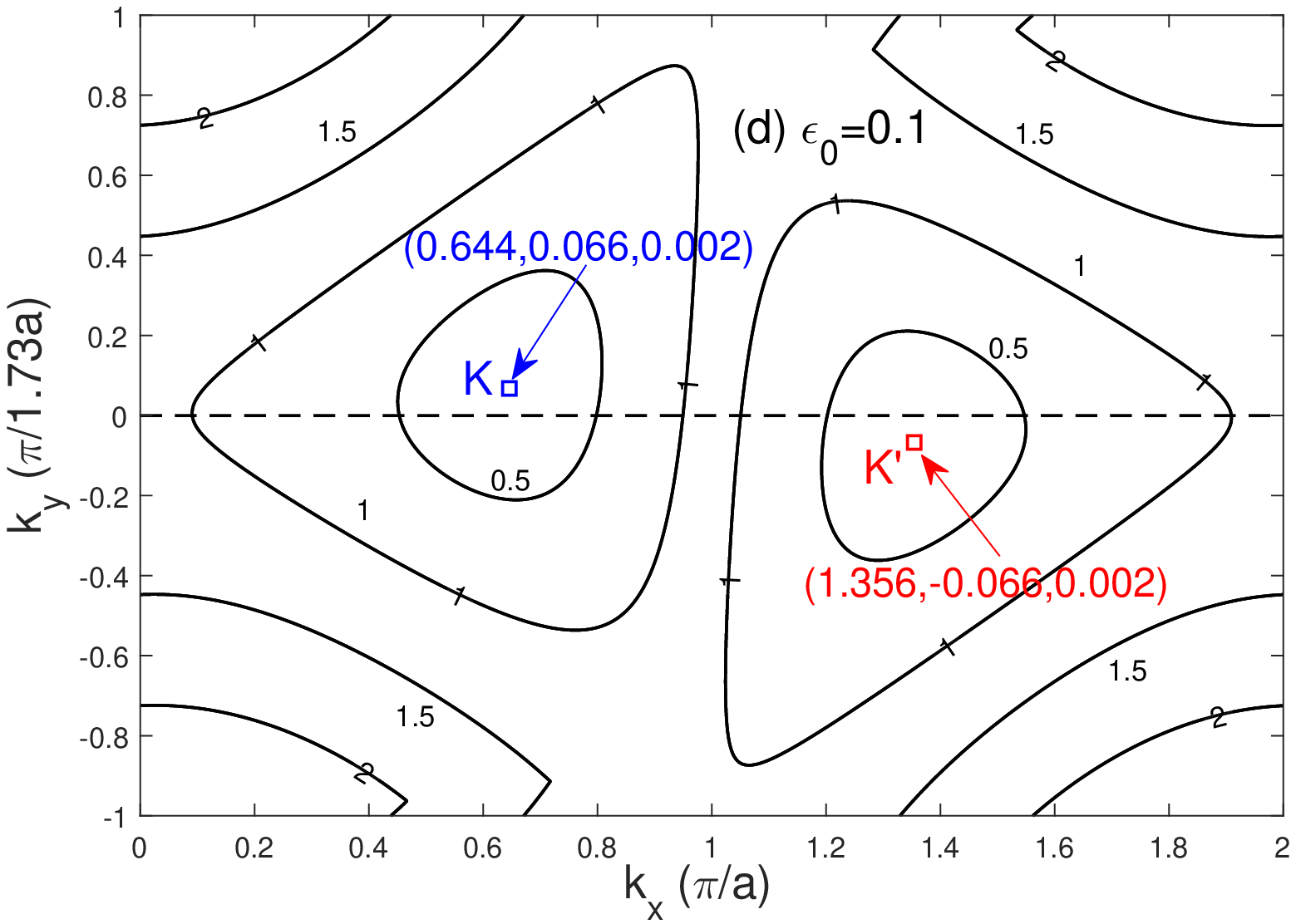}
\caption{\label{fig:Coutour} Contour plot of the dispersion relations as a function of the wave vectors $k_x$
and $k_y$ for different strain strengths: (a) $\epsilon_0=0$, (b) $\epsilon_0=0.02$, (c) $\epsilon_0=0.05$ and (d) $\epsilon_0=0.1$.
The solid lines are equal energy contour lines corresponding to energy values of $0.5 \mathrm{eV}$, $1 \mathrm{eV}$, $1.5 \mathrm{eV}$,
and $2 \mathrm{eV}$.
The parameters are $E_z=16.96 \mathrm{meV}{\AA}$, $\nu_{\parallel}=0.25$, $\nu_{\perp}=2.5$, and $\phi=30^\circ$.
The blue and red squares refer to the Dirac points of K and K' with the value of $(k_x, k_y, E)$.}
\end{figure*}
In the central scattering region, the silicene sheet is uniformly stretched (or compressed) along the angle $\phi$
relative to the axis $x$. Note that we assume there exists no strain outside the central scattering region.
In the considered Cartesian coordinates,
the tension $\mathbf{T}$ can be written as $\mathbf{T}=\mathrm{T}(\cos\phi \hat{e}_x+\sin\phi \hat{e}_y)$.
It is convenient to represent the tension in the principal coordinates $Ox'y'$, i.e., $\mathbf{T}=\mathrm{T}\hat{e}_{x'}$.
In terms of the generalized Hooke's law~\cite{Pereira2}, the strain $\epsilon'_{ij}$ are related to the components of the compliance tensor, namely
$\epsilon'_{ij}=\mathrm{T}S_{ijxx}$,
with the indices $i, j=x,y,z$. For the honeycomb lattice, we know that only five compliance tensor components are independent
(i.e., $S_{xxxx}$, $S_{xxyy}$, $S_{xxzz}$, $S_{zzzz}$, $S_{yzyz}$)~\cite{Blakslee}.  Thus, the Poisson's
transverse ratio and perpendicular ratio are defined as
\begin{eqnarray}
\nu_{\parallel}=-S_{xxyy}/S_{xxxx}, \nu_{\perp}=-S_{xxzz}/S_{xxxx}.
\end{eqnarray}

When the strain is applied to the low-buckled geometry, the lattice deformation will result in the
change of the vectors $\xi_\ell$ ($\ell=1,2,3$). Expanded in the first-order approximation,
the strain-dependent vectors are given by $\vec{\xi}_{\ell}=(1+\vec{\epsilon})\vec{\xi}_{\ell}^0$, which thus
modulates the hopping terms. Accordingly, we obtain the deformed bond length as follows
\begin{eqnarray}
|\vec{\xi}_{1}|&=&\big\{\frac{\sqrt{2}}{10}\epsilon_0[(2\cos^2\phi+\sqrt{3}\sin2\phi)(1+\nu_{\parallel})\\
&&+(1-3\nu_{\parallel}-\frac{\nu_{\perp}}{6})]+\frac{5\sqrt{2}}{12}\big\}a,\nonumber\\
|\vec{\xi}_{2}|&=&\big\{\frac{\sqrt{2}}{10}\epsilon_0[(2\cos^2\phi-\sqrt{3}\sin2\phi)(1+\nu_{\parallel})\nonumber\\
&&+(1-3\nu_{\parallel}-\frac{\nu_{\perp}}{6})]+\frac{5\sqrt{2}}{12}\big\}a,\nonumber\\
|\vec{\xi}_{3}|&=&\big\{\frac{2\sqrt{2}}{5}\epsilon_0[(1-\frac{\nu_{\perp}}{24})-\cos^2\phi(1+\nu_{\parallel})]
+\frac{5\sqrt{2}}{12}\big\}a.\nonumber
\end{eqnarray}

The height is $h=\sqrt{2}(1-\nu_\perp\epsilon_0)/12$. As the deformation is increased to $\epsilon_0=1/\nu_\perp$, the buckled structure is
gradually stretched to a planar structure.

For the low-buckled silicene described by $\mathrm{s}$ and $\mathrm{p}$ orbitals,
there are four types of hopping integrals $V_{ss\sigma}$, $V_{sp\sigma}$, $V_{pp\sigma}$ and $V_{pp\pi}$.
Within the Slater-Koster framework~\cite{Slater}, the hopping processes between the two neighboring sites depend only
on the bond length and the relative angle $\Omega$.
The hopping parameters in equation (\ref{eq:parameter}) can be calculated in terms of the formula given
in Ref. \cite{Liu}, which considered the weak contribution of the angle $\Omega$ on the hopping processes.
Under the two-center approximation adopted by Slater and Koster, the hopping integrals can be expressed as~\cite{Yip}
\begin{eqnarray}
V_\mu(r_{\ell})=\alpha_1r_{\ell}^{-\alpha_2}\exp(-\alpha_3r_{\ell}^{\alpha_4}),
\end{eqnarray}
where $\mu$ refers to the four types of the hopping integrals, $r_{\ell}=|\vec{\xi}_{\ell}|$ is the bond length,
$\alpha_{\kappa}(\kappa=1, 2, 3, 4)$ denotes the system parameters for silicene. So far, there are no microscopic evaluations of the
four parameters for the silicene sheet from experiments and first-principle calculations. We slightly modify the parameters obtained
from Environment-dependent tight-binding potential model~\cite{Yip} to mimic the hopping integral of the silicene sheet.

The Possion's transverse ratio and perpendicular ratio can change slightly with increasing strain~\cite{Wang}.
For simplicity, in our numerical calculation, we choose the Possion's ratio to be $\nu_{\parallel}=0.25$ and $\nu_{\perp}=2.5$. For this case,
the low-buckled silicene will be stretched to a planar structure when the strain is close to $0.4$.

We first investigate the dispersion relation of the infinite-sized, homogeneous silicene sheet under the influence of
the strain $\epsilon_0$ and the electric field $E_z$, as shown in Fig.~\ref{fig:dispersion}.
In the absence of the strain and the electric field, the energy band is spin and valley-degenerate and has a small band gap of about $8.2 \mathrm{meV}$ arising from the effective spin-orbit coupling.
When the electric field is increased to the critical value $E_{zc}=t_{so}/a_z=16.96 \mathrm{meV}{\AA}$
[see Fig.~\ref{fig:dispersion}(b)], we find that the band gap gradually approaches to zero for up-spin electrons
at K valley and down-spin electrons at K' valley. Correspondingly, the spin and valley-degeneracy are broken
by the electric field. Thus the electrons can become perfectly spin-up (spin-down) polarized at the K (K') point under the influence of the electric field $E_{zc}$, which agrees well with
the results obtained from the low-energy theory~\cite{Ezawa2}. When the strain is applied to the silicene system,
for example when $\epsilon_0=0.005$, the energy difference of the conduction band and the valence band increases to $5.6 \mathrm{meV}$.
With increasing amplitude of the strain to $\epsilon_0=0.05$, as shown in Fig.~\ref{fig:dispersion}(d),
the energy difference is significantly enlarged to about $100 \mathrm{meV}$ for the two valleys.
Especially, the spin-polarization induced by the electric field is also suppressed, and the dispersion relation recovers
the spin and valley degeneracy.

It is natural to consider whether the minima of energy profile for the two valleys still concide at $k_y=0$ in the presence of the strain. In order to clarify the effect of the strain on Dirac points, we plot the dispersion relations as a function of the wave vectors $k_x$ and $k_y$ for different strain strengths, as shown in Fig.~\ref{fig:Coutour}. We can see that, when $\epsilon_0=0$, the Dirac points of K and K' are located at the points with the wave vectors $(k_x,k_y)=(0.667,0)$ and $(1.333,0)$, respectively. Interestingly, the Dirac point K moves towards the positive direction of $k_y$ axis, while the Dirac point K' moves towards the negative direction with the strain increasing from $\epsilon_0=0$ to $0.1$. Thus, the application of strain results in a relative transverse shift of the two Dirac cones. At the same time, the Dirac points of K and K' also move away from each other along the $k_x$ direction. Therefore, the dispersion relations in Fig.~\ref{fig:dispersion}(c) and (d) are just representing a cut of the Dirac cone at $k_y=0$, and that the energy difference between the conduction and valence bands depicted there are not the actual band gap. 
For comparison, we recall the graphene system where the strain can induce pseudo-magnetic fields greater than $300$ Tesla~\cite{Levy}.
This pseudo-magnetic field can be described by a gauge field $\mathcal{A}$ in the low-energy approximation~\cite{Fogler}.
Correspondingly, the Hamiltonian of the strained graphene sheet has the form~\cite{Pereira,Fujita}
$H=v_F\sigma\cdot(\vec{p}-\mathcal{A}/v_F)$, where $\mathcal{A}$ has reversed signs for valleys K and K'.
Considering the strain-induced deflection of the Dirac points of K and K', it is reasonable to deduce that one can also adopt a gauge field to effectively describe the strain effect in the silicene sheet.

\begin{figure}[b]
\includegraphics[width=7.0cm]{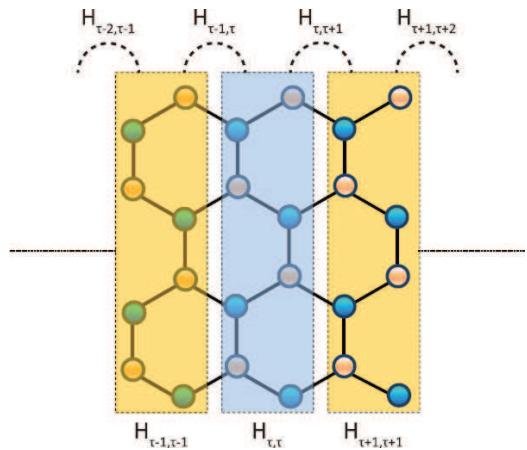}
\caption{\label{fig:device2} The system is divided into cells indicated by an index $\tau$.
$H_{\tau,\tau}$ is the Hamiltonian matrix representing the hopping terms between sites within cell $\tau$;
$H_{\tau, \tau\pm 1}$ is the Hamiltonian matrix connecting the sites between neighboring cells.}
\end{figure}
\section{CALCULATION OF TRANSPORT PROPERTY}\label{sec:transport}
In order to calculate the transport property, we adopt the method formulated by Ando~\cite{Ando}.
The silicene heterojunction is divided into cells indicated by an index $\tau$, which represents
a minimum repeating unit, as shown in Fig.~\ref{fig:device2}. The source and drain are
ideal leads that span the cells $\tau=-\infty,\ldots, 0$ and $\tau=S+1,\ldots, \infty$. The central scattering region
spans the cells $\tau=1,2,\ldots, S$. The Schr\"{o}dinger equation of the silicene heterojunction can be written as
\begin{eqnarray}\label{eq:Bloch}
-\mathbf{H}_{\tau,\tau-1}\mathbf{\psi}_{\tau-1}+(E\mathbf{I}-\mathbf{H}_{\tau,\tau})
\mathbf{\psi}_\tau-\mathbf{H}_{\tau,\tau+1}\mathbf{\psi}_{\tau+1}=0,
\end{eqnarray}
for $\tau=-\infty,\ldots, \infty$. If each cells contains $N$ orbitals, $\mathbf{\psi}_\tau$ is a $N$ dimensional
vector including the wave-function coefficients of all orbitals for cell $\tau$.
$\mathbf{H}_{\tau,\tau}$ is the $N\times N$ Hamiltonian matrix representing the hopping terms between sites within cell $\tau$,
$\mathbf{H}_{\tau, \tau\pm 1}$ is the $N\times N$ Hamiltonian matrix connecting the sites between neighboring cells, which can be mapped from
the tight-binding Hamiltonian in Eq.~(\ref{eq:parameter}).

First, we need to find the solutions of the wavefunction in leads. Since the leads have periodic structures, the vectors in subsequent cells satisfy the Bloch condition
\begin{eqnarray}
\mathbf{\psi}_\tau=\lambda \mathbf{\psi}_{\tau-1},
\end{eqnarray}
where $\lambda=e^{ik_xa}$ is the Bloch factor with $k_x$ real for propagating waves
and complex for evanescent waves. Substituting this formula into Eq.~(\ref{eq:Bloch}) for left and right leads,
we can obtain the generalized eigenvalue equation:
\begin{widetext}
\begin{eqnarray}
 \Big[\left(
\begin{array}{ccc}
E\mathbf{I}-\mathbf{H}_{\tau,\tau} & \mathbf{H}_{\tau,\tau+1} \\
\mathbf{I} & 0
 \end{array}
 \right)-\lambda\left(
\begin{array}{ccc}
-\mathbf{H}^\dag_{\tau,\tau+1} & 0 \\
0 &\mathbf{I}
 \end{array}
 \right)\Big]\left(
\begin{array}{ccc}
\mathbf{\psi}_\tau\\
\mathbf{\psi}_{\tau-1}
 \end{array}
 \right)=0
\end{eqnarray}
\end{widetext}

After solving this equation, we obtain nontrivial solutions, which can be divided into propagating modes
and evanescent modes in terms of the eigenvalues~\cite{Khomyakov}. The eigenvalues of the propagating modes and the evanescent modes
satisfy the conditions $|\lambda(\pm)|=1$ and $|\lambda(\pm)|\neq 1$, respectively, with $+/-$ referring to the right-going modes and left-going modes. When $|\lambda(+)|< 1$, the eigenvector is named as right-going evanescent modes,
while the states with $|\lambda(-)|> 1$ associate with left-going evanescent modes. For the propagating states, in order to distinguish
the right- and left-going modes, one needs to calculate their Bloch velocities
\begin{eqnarray}
v_n(\pm)=-\frac{2a}{\hbar}\mathrm{Im}[\lambda_n(\pm)\psi_n(\pm)^\dag H^\dag_{\tau,\tau+1}\psi_n(\pm)],
\end{eqnarray}
where the sign of the velocities distinguishes right from left propagation.
Accordingly, we can distinguish the valleys K and K' in terms of the wave vector
$k_x$ derived from the eigenvalue $\lambda$. The first valley K is related to the longitudinal wave vector $k_xa\in (0, \pi)$,
whereas the second valley K' lies in the wave vector regime $k_xa\in (\pi, 2\pi)$~\cite{Rycerz}.

The general solution of the leads can be written as
\begin{eqnarray}
\psi_\tau &=&\psi_\tau(+)+\psi_\tau(-)\nonumber\\
&=&\mathbf{F}^{\tau-\tau'}(+)\psi_{\tau'}(+)+\mathbf{F}^{\tau-\tau'}(-)\psi_{\tau'}(-),
\end{eqnarray}
where the matrices $\mathbf{F}(\pm)$ are defined as
\begin{eqnarray}
\mathbf{F}(\pm)=\sum_n^N\lambda_n(\pm)\psi_n(\pm)\tilde{\psi}^\dag_n(\pm),
\end{eqnarray}
where $\tilde{\psi}_n(\pm)$ are dual vectors, which satisfy the following relations
\begin{eqnarray}
\tilde{\psi}^\dag_n(\pm)\psi_m(\pm)=\delta_{n,m}, \psi^\dag_n(\pm)\tilde{\psi}_m(\pm)=\delta_{n,m}.
\end{eqnarray}
Note that the eigenvectors are nonorthogonal.

Next we calculate the solutions of the scattering regions.
By treating the effect of the leads as the boundary conditions,
the Schr\"{o}dinger equation of the scattering region can be modified as
\begin{eqnarray}\label{eq:Hamiltonian}
&&-\mathbf{H}'_{\tau,\tau-1}\psi_{\tau-1}+(E\mathbf{I}-\mathbf{H}'_{\tau,\tau})\psi_{\tau}-\mathbf{H}'_{\tau,\tau+1}\psi_{\tau+1}\nonumber\\
&&=\mathbf{\Lambda}_0\psi_0(+)\delta_{\tau,0},
\end{eqnarray}
where the index of the cells becomes $\tau=0,1,,\ldots, S,S+1$.
The renormalized Hamiltonian matrices are
\begin{eqnarray}
\mathbf{H}'_{0,0}=\mathbf{H}^L_{\tau',\tau'}+\mathbf{H}^L_{\tau',\tau'+1}\mathbf{F}^{-1}_L(-),\nonumber\\
\mathbf{H}'_{S+1,S+1}=\mathbf{H}^R_{\tau',\tau'}+\mathbf{H}^{R\dag}_{\tau',\tau'+1}\mathbf{F}^{-1}_R(+),\nonumber\\
\mathbf{H}'_{0,-1}=0,\mathbf{H}'_{S+1,S+2}=0,
\end{eqnarray}
and other Hamiltonian matrices are $\mathbf{H}'_{\tau,\tau'}=\mathbf{H}_{\tau,\tau'}$ for the indexes $\tau,\tau'=0,1,\ldots, S,S+1$.
The source term is $\mathbf{\Lambda}_0=\mathbf{H}^L_{\tau',\tau'+1}[\mathbf{F}^{-1}_L(+)-\mathbf{F}^{-1}_L(-)]$
with $L/R$ referring to the left and right leads.
\begin{figure}[t]
\includegraphics[width=4cm]{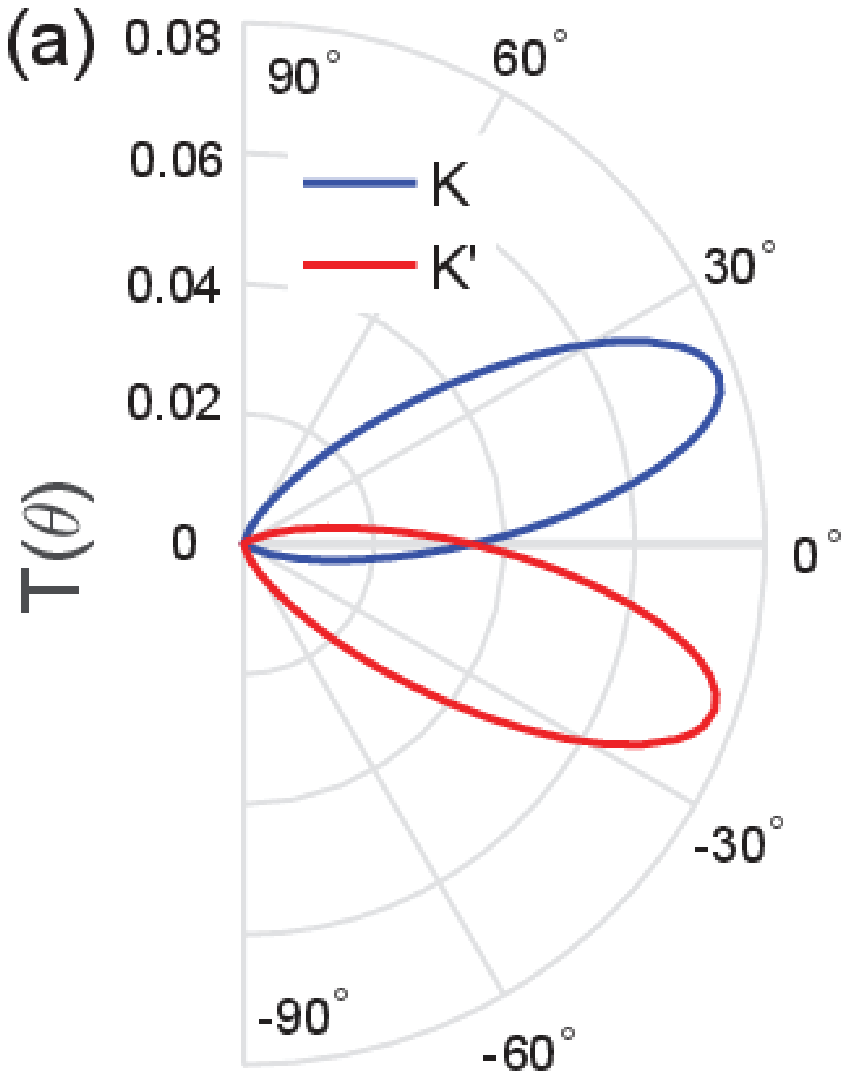}
\includegraphics[width=4cm]{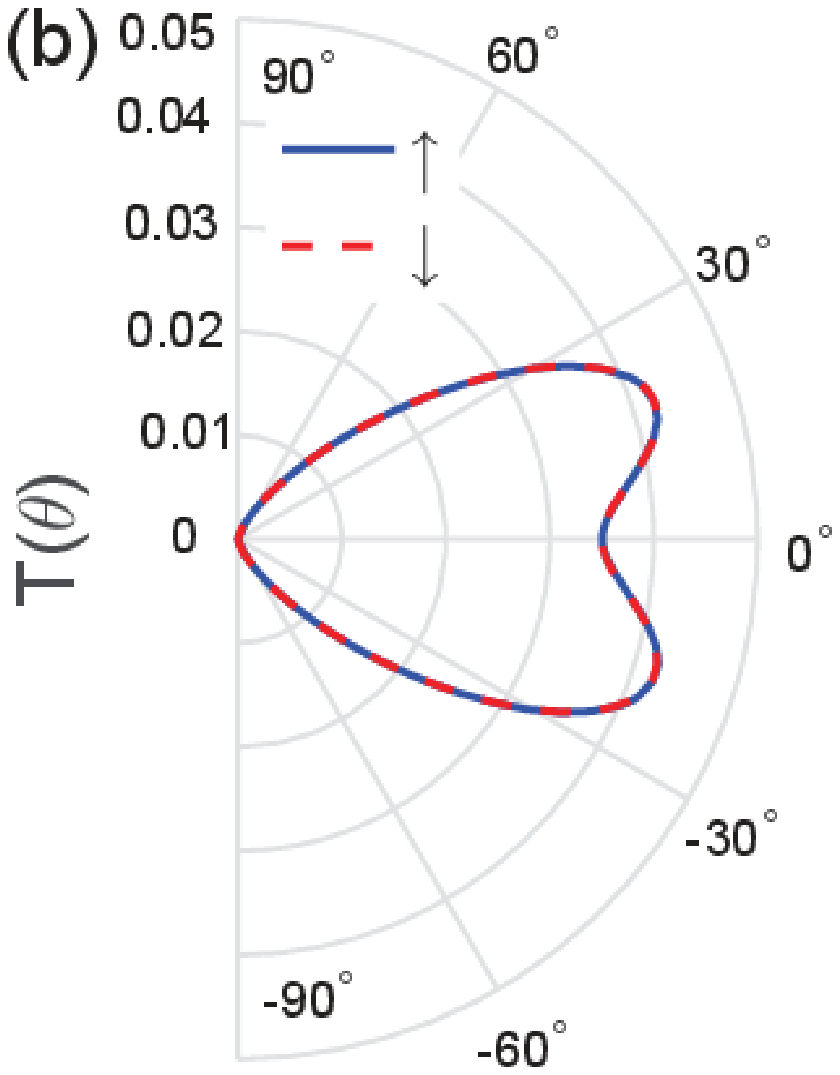}
\caption{\label{fig:TransT5} (a) The valley dependence and (b) spin dependence of the transmission are plotted as
a function of the incident angle $\theta$ for $E_z=0$ and $\epsilon_0=0.005$.
The other parameters are $E_F=7.9 \mathrm{meV}$, $V_0=8.4\mathrm{meV}$ and $L=193\mathrm{nm}$.}
\end{figure}

Eq.~(\ref{eq:Hamiltonian}) gives a set of linear equations, which can be solved efficiently by using the block Gaussian elimination method. We can then obtain the transmission matrix elements $t_{n,m}$ by expanding the vector $\psi_{S+1}(+)$ in modes of the right lead
\begin{eqnarray}
\psi_{S+1}(+)=\sum_n^N \psi_{R,n}(+)t_{n,m},
\end{eqnarray}
where the incoming wave is chosen as one of the propagating modes of the left lead, namely $\psi_0(+)=\psi_{L,m}$. After running the vector $\psi_0(+)$ runs over all possible modes of the left lead, namely $\psi_{L,m}, m=1,2,\ldots,N$, the full transmission matrix can be obtained.

Accordingly, the total transmission can be written as
\begin{eqnarray}
T_{LR}(k_y,E)=\sum_{n,m}^N \frac{v_{R,n}}{v_{L,m}}|t_{n,m}|^2,\nonumber\\
t_{n,m}=\tilde{\mathbf{\psi}}^\dag_{R,n}(+)G_{S+1,0}[G_{0,0}^{(0)}]^{-1}\psi_{L,m}(+),
\end{eqnarray}
where $G_{0,0}^{(0)}$ and $G_{S+1,0}$
refer to the Green's function of the left lead and the full system, respectively, which can be
obtained by using the iterative techniques of Green's function approach~\cite{Khomyakov}.
After obtaining the Green's functions, we can calculate the valley-resolved transmission in terms of the corresponding eigenvalues.

Utilizing the periodical boundary conditions at the transverse direction, we can introduce the wave vector $k_y$
into the Hamiltonian and effectively mimic the silicene sheet by using a silicene nanoribbon with zigzag chain number of $N_y=2$~\cite{Li2}.
The incident angle is defined as
$\theta=\arcsin(k_y/k_F)$, where the Fermi wave vector $k_F$ can be obtained from the relation~\cite{Ezawa2}
\begin{eqnarray}
E_F=\sqrt{\hbar^2v^2_Fk^2_F+\Big(a_zE_z-\sqrt{t_{so}^2+a^2t_{R2}^2k_F^2}\Big)^2},
\end{eqnarray}
with $v_F=\sqrt{3}a t/2$ being the Fermi velocity.

\begin{figure}[t]
\includegraphics[width=4.2cm]{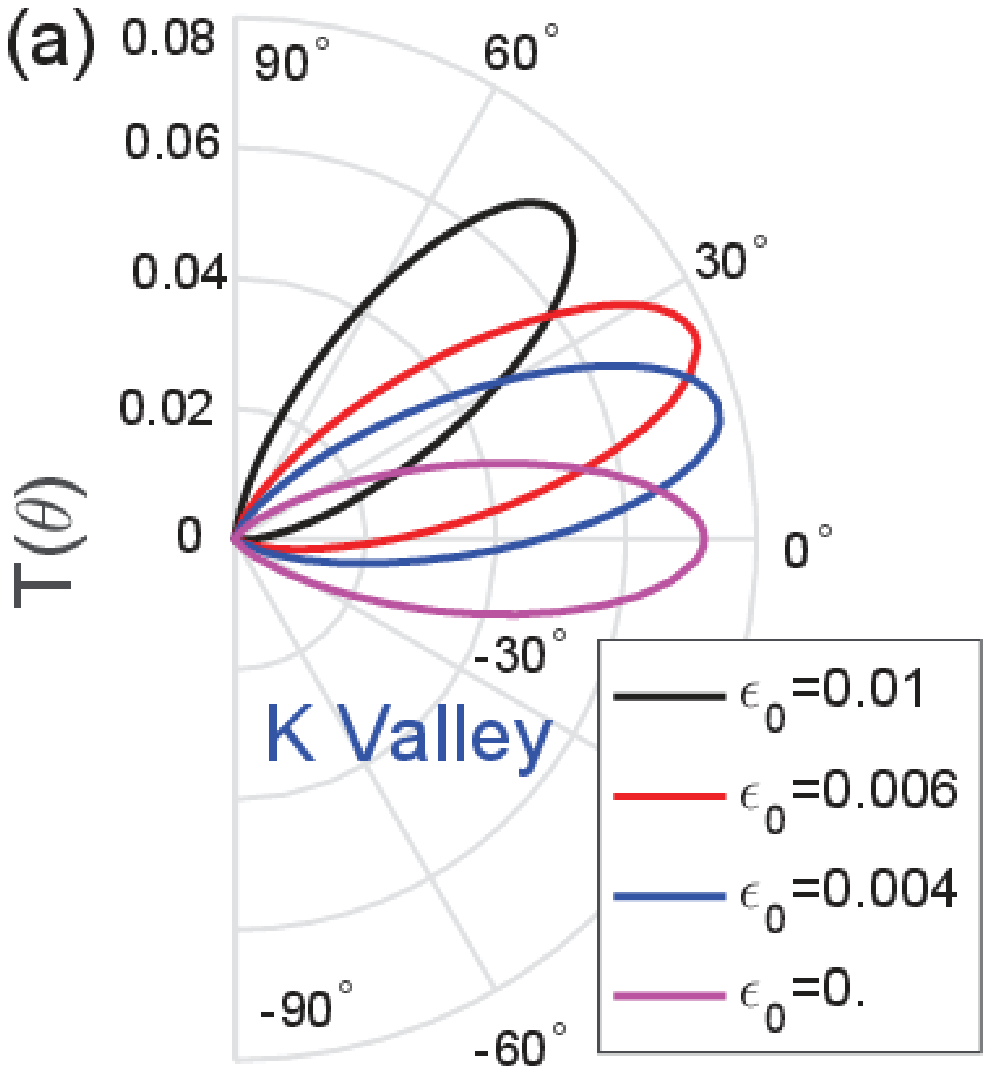}
\includegraphics[width=4.2cm]{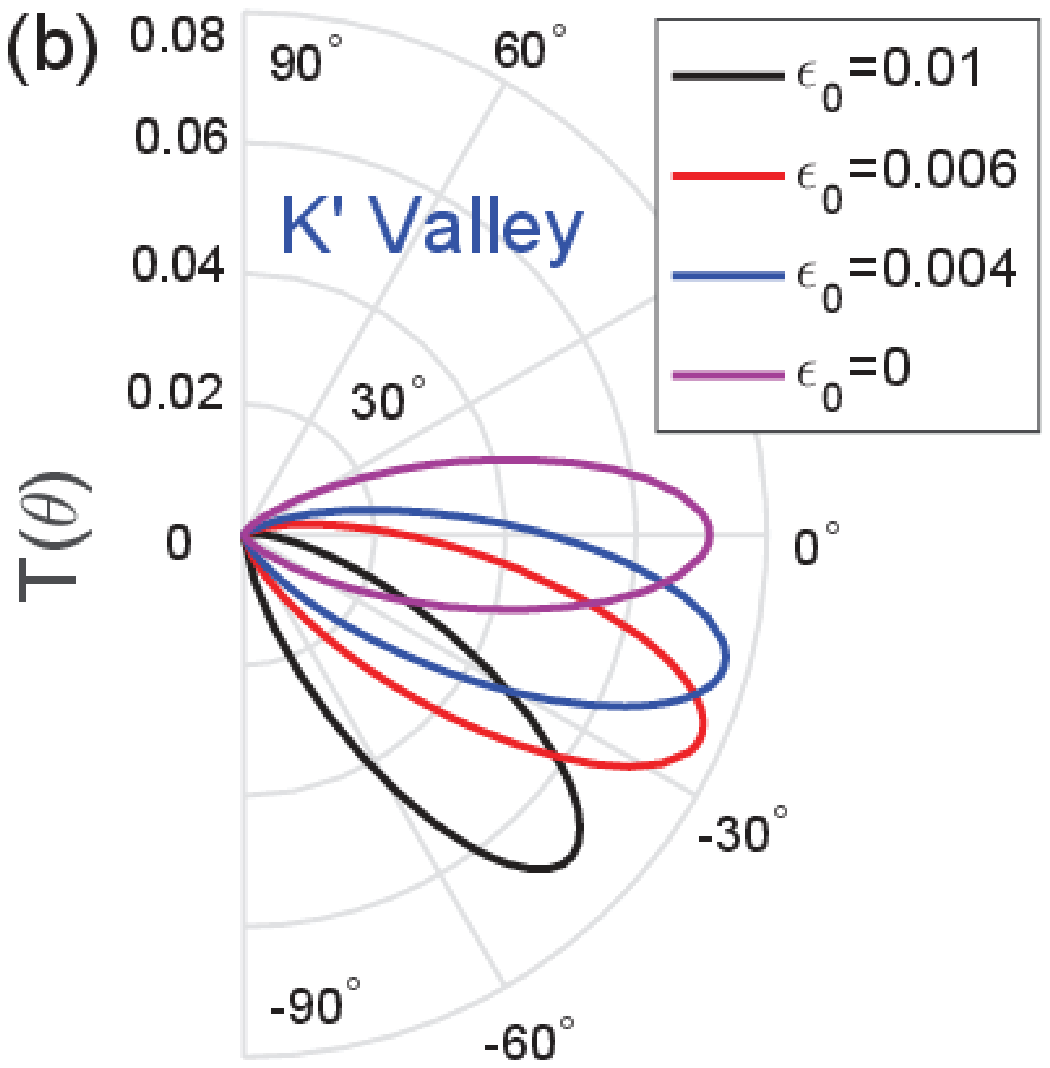}
\caption{\label{fig:TransValley} (a)The K valley (b) and K' valley dependence of the transmission are plotted as
a function of the incident angle $\theta$ for different amplitudes of the strain.
The other parameters are $E_z=0$, $E_F=7.9 \mathrm{meV}$, $V_0=8.4\mathrm{meV}$ and $L=193\mathrm{nm}$.}
\end{figure}

\section{RESULTS AND DISCUSSION}\label{sec:Results}
It is found that the valley-dependent and spin-dependent electrons cannot be dispersed by only the electric field.
We thus consider the effect of the strain on the transport properties.
In the central scattering region, the
silicene sheet is uniformly stretched along the angle $\phi=30^\circ$ relative to the axis x with $E_z=0$. When the strain is $\epsilon_0=0.005$,the transmission curve of K valley is deflected upwards, while the curve of K' valley is deflected downwards,
as shown in Fig.~\ref{fig:TransT5}(a). Moreover, the maximum value of the transmission is significantly reduced to about $0.08$ due to the effect of the strain.
It means that the strain can result in the separation of Dirac fermions of K and K' valleys,
which is similar with the deflection behavior induced by real magnetic fields in graphene systems~\cite{Martino,Masir,Li3}.

In Fig.~\ref{fig:TransT5}(b), we see that the spin-dependent transmission is decreased to about $0.052$.
However, the transmission profiles of up-spin and down-spin electrons are identical and symmetric with respect to normal incidence.Thus, the strain can separate the electrons of valleys K and K' but cannot separate the up-spin and down-spin electrons.

\begin{figure}[t]
\includegraphics[width=3.8cm]{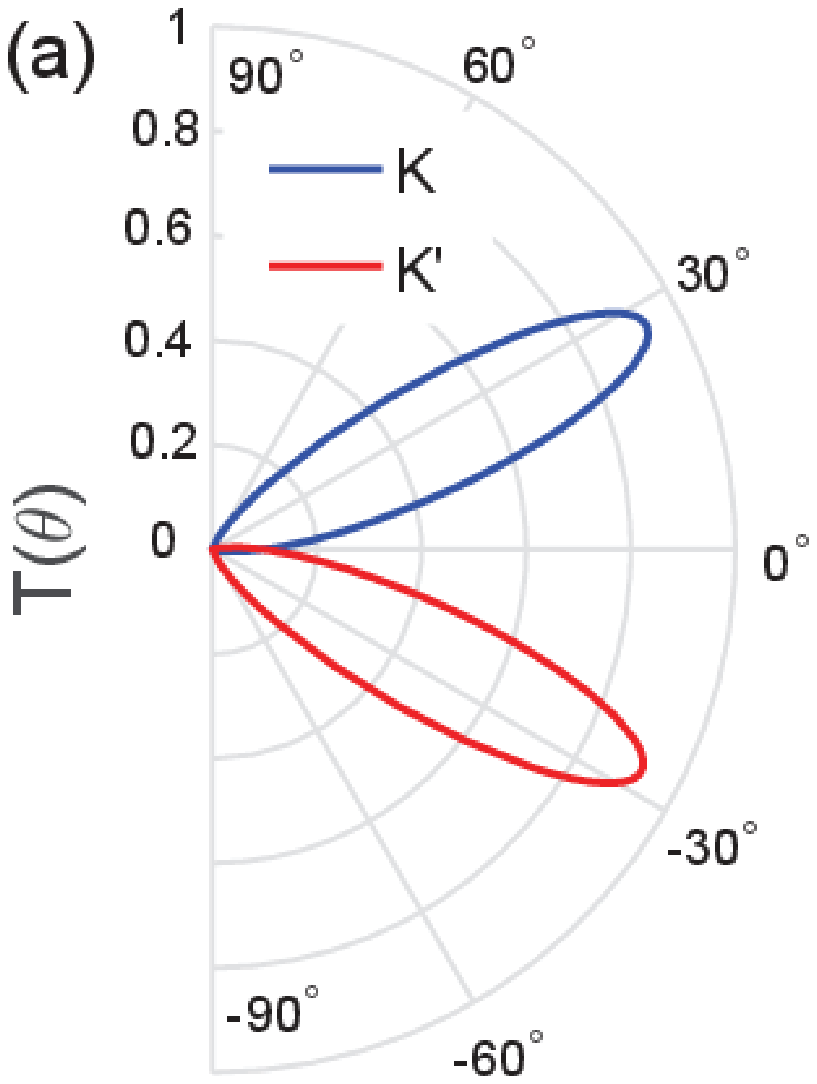}
\includegraphics[width=3.8cm]{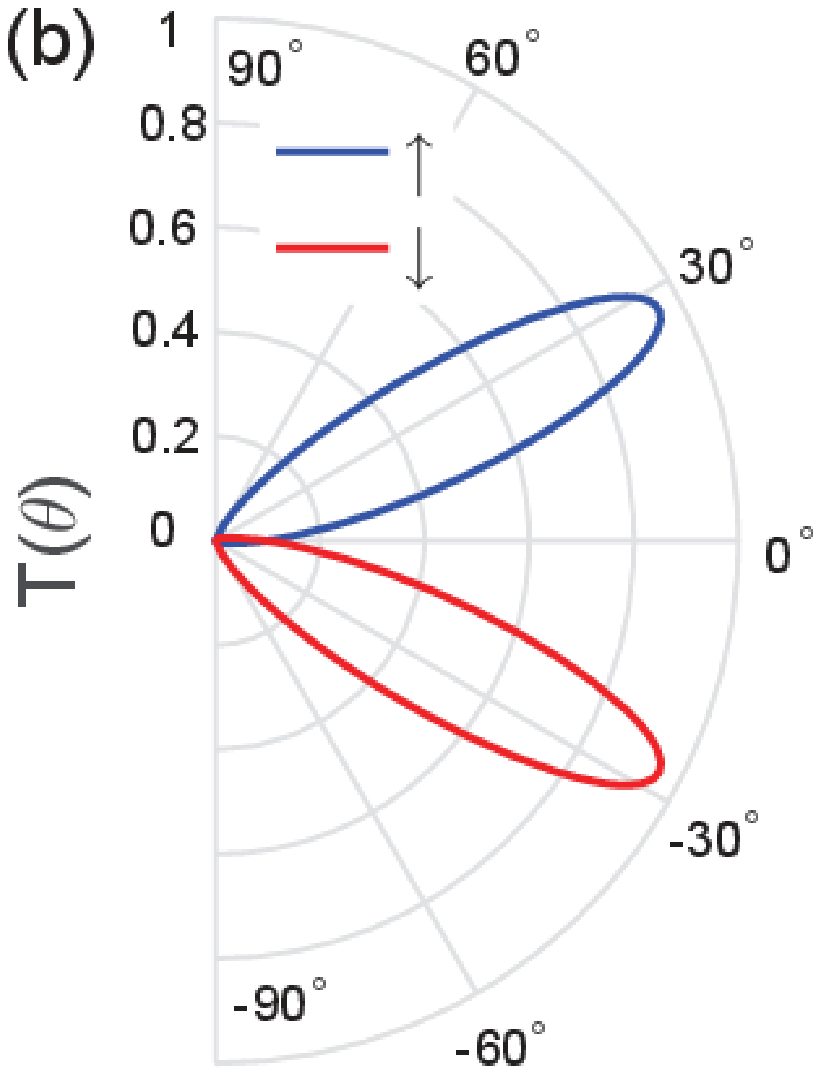}
\caption{\label{fig:TransSpin} (a) The valley dependence and (b) spin dependence of the transmission are plotted as
a function of the incident angle $\theta$ for $E_{z}=16.96 \mathrm{meV}{\AA}$ and $\epsilon_0=0.006$.
The other parameters are $E_F=7.9 \mathrm{meV}$, $V_0=8.4\mathrm{meV}$ and $L=193\mathrm{nm}$.}
\end{figure}

In order to clarify the effect of the strain on the valley-dependent transport, we plot the transmission of valleys K and K'
as a function of the incident angle $\theta$ for different amplitudes of the strain, as shown in Fig.~\ref{fig:TransValley}.
We find that the transmission profile of K valley is deflected upwards with increasing strain from $0.004$ to
$0.01$. When $\epsilon_0=0.01$, the transmission of electrons is pushed towards the angular regime
$\theta>45^\circ$ [see Fig.~\ref{fig:TransValley}(a)]. Seen from the physical picture of view,
the K-valley electrons will be scattered back the left region
if the incident angle is smaller than a certain critical angle.
Correspondingly, the transmission profile of the K' valley is deflected downwards under the influence of the strain.
The K'-valley electrons will be deflected back into the incident region when the incident angle is larger than a certain angle.
When the strain is along the zigzag or armchair direction, namely $\phi=0^\circ$, $\phi=90^\circ$
and $\phi=-90^\circ$, the transmission profiles have no deflection behavior, which is a distinct anisotropy behavior for
the strain modulation of the valley current.

The above analysis show that strain can be utilized to separate the Dirac fermions of different valleys. Since silicene has a spin-valley correlation, it is thus natural to think that we can separate the electrons of different spins in the silicene sheet by applying the strain and an external electric field. Fig.~\ref{fig:TransSpin} gives a clear picture of the strain modulation of spin and valley components.
When $E_{z}=16.96 \mathrm{meV}{\AA}$ and $\epsilon_0=0.006$, the transmission curves of valleys K and K' still deflect
upwards and downwards, respectively[see Fig.~\ref{fig:TransSpin}(a)].
The transmission profile of up-spin (down-spin) electrons is also obviously pushed upwards (downwards). This
shows that the up-spin (down-spin) component is related to K (K') valley. Therefore, one can separate
the electrons of different spins into two opposite transverse directions, which can result in a strain-induced spin (valley) Hall effect in
a suitable silicene device. This phenomenon is similar with the spin-valley Hall effect reported in monolayer graphene~\cite{Islam}.
\begin{figure}[t]
\includegraphics[width=4.2cm]{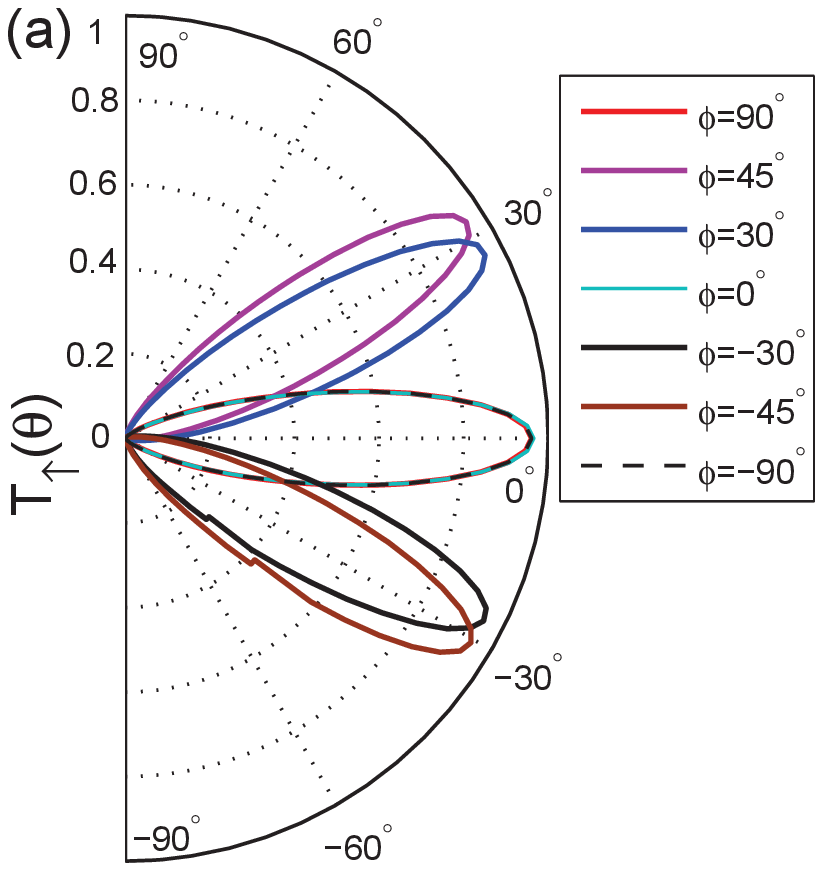}
\includegraphics[width=4.2cm]{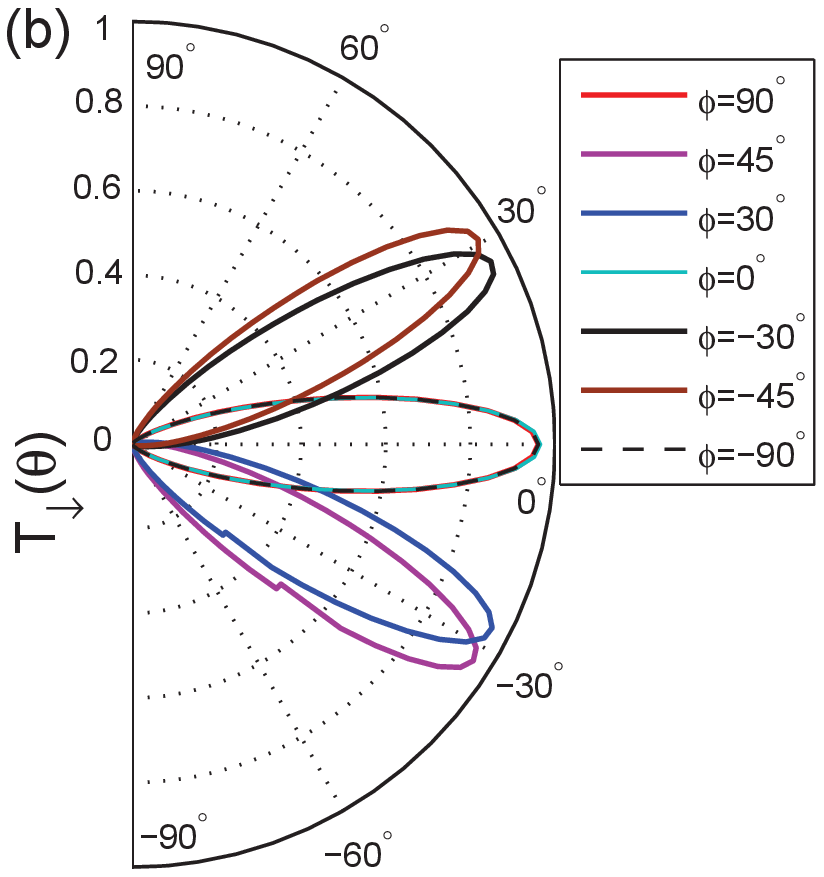}
\caption{\label{fig:TransSpinPhi} The transmission of (a) up-spin and (b) down-spin electrons are plotted as
a function of the incident angle $\theta$ for different angles $\phi$.
The other parameters are $E_{z}=16.96 \mathrm{meV}{\AA}$, $\epsilon_0=0.006$, $E_F=7.9 \mathrm{meV}$,
$V_0=8.4\mathrm{meV}$ and $L=193\mathrm{nm}$.}
\end{figure}

Similarly, the up-spin (down-spin) transmission curves are deflected upwards (downwards) when $\phi=30^\circ$ and $\phi=45^\circ$,
as shown in Fig.~\ref{fig:TransSpinPhi}.
However, when the angle is changed to negative values, namely $\phi=-30^\circ$ and $\phi=-45^\circ$, the transmission profiles
of up-spin and down-spin components are pushed downwards and upwards, respectively.
When the strain is along the zigzag or armchair direction, the transmission profiles of two spin
components are symmetrical with respective to the normal incident, so the up-spin and down-spin electrons can not be separated
at these strain configurations. These results imply that one can modulate the spin polarization
by changing the stretching angle of the strain.

Since germanene also has a honeycomb geometry~\cite{Liu2,Ezawa} and its Hamiltonian is the same as equation~(\ref{eq:parameter}), germanene can be modeled by replacing the parameters with $t=1.3 \mathrm{eV}$, $t_{so}=43 \mathrm{meV}$, $t_{R2}=10.7 \mathrm{meV}$ and $a_z=0.33 {\AA}$. The band gap induced by the spin-orbit couplings can reach $93 \mathrm{meV}$~\cite{Liu}, which can provide a significant modulation of spin- and valley-dependent properties. We think one can also observe the spin and valley separation in germanene systems due to its similar geometry and low-buckling structure. The numerical trends of the spin-valley separation due to strain in germanene systems would be the same as those shown in Figs.~\ref{fig:TransT5}-\ref{fig:TransSpinPhi} for silicene.

However, the parameters (e.g. the change in the bond length, and the $\alpha$ coefficients in the Slater-Koster integral) under the influence of strain and the electric field would be different. Their exact values need to be determined e.g. by ab initio calculations, and currently they are not available in the literature, unlike for silicene. However, given the larger SOC values in germanene, we believe that one would require a relatively smaller amplitude of the strain to realize the same degree of valley and spin separation in comparison with silicene systems. Thus, in experimental investigation of germanene systems, we envisage that would be easier to realize the valley and spin separation by applying similar strain and the electric field configurations as that assumed in this paper for the strained silicene system. This suggests
that the strain-induced valley and spin separation can in general be observed in 2D materials with low-buckled honeycomb structures.

\section{CONCLUSIONS}\label{sec:Conclusions}
In summary, we have studied the effect of the strain and the external electric field on the
dispersion relation and the transport property of a silicene heterojunction. It is found that the valley-dependent
and spin-dependent electrons cannot be dispersed only by the electric field. In the presence of the strain,
the transmission profiles can be deflected to two opposite transverse directions, thus resulting in the separation
of valleys K and K'. When a strain and an electric field are  applied to the scattering region simultaneously, not only are the electrons of valleys K and K' separated into two branches,
but the up-spin and down-spin electrons will also move towards two opposite transverse directions correspondingly. Therefore, combining the strain and the electric field, one can realize an effective modulation of the valley-dependent and spin-dependent transport by changing the amplitude and the stretching direction of the strain.
Our results may be helpful for exploring the transport mechanism of strain modulated silicene systems and making
the new types of the silicene-based valleytronics and spintronics devices.

\acknowledgments This work was supported by NSF-China (Grant No. 11574067), Natural Science Foundation of Zhejiang Province (Grant Nos. LY16A040007, LY15E060007, LQ17F010004), and the National Research Foundation of Singapore under the Competitive Research Program
 "Non-Volatile Magnetic Logic And Memory Integrated Circuit Devices" NRF-CRP9-2011-01.

\end{document}